\documentclass[%
 reprint,
 amsmath,amssymb,
 aps,
 pra,
]{revtex4-1}

\usepackage{graphicx}
\usepackage{dcolumn}
\usepackage{bm}
\usepackage{subfig}
\usepackage{amsmath}

\begin{document}

\title{Non-linear impedance spectroscopy applied to thermoelectric measurements: beyond the zT estimation.}

\author{E. Thi\'ebaut$^1$}

\author{F. Pesty$^1$}

\author{C. Goupil$^2$}

\author{G. Guegan$^3$}

\author{P. Lecoeur$^1$}

\affiliation{$^1$Centre de Nanosciences et de Nanotechnologies, CNRS, Univ. Paris-Sud, Universit\'e Paris-Saclay, C2N-Orsay, 91405 Orsay, France \\
$^2$Laboratoire Interdisciplinaire des Energies de Demain (LIED), UMR 8236 Universit\'e Paris Diderot, CNRS, 5 Rue Thomas Mann, 75013 Paris, France \\
$^3$STMicroelectronics TOURS, 10 Rue Thales de Milet, CS97155, 37071 Tours Cedex 2, France}

\begin{abstract}

Thermoelectric measurement of the dimensionless $zT$ parameter requires multiple physical quantities to be measured, therefore there is great interest to find an experimental setup capable of measuring all these properties at once. 
Previous works on impedance spectroscopy have shown promising results in this direction, however, this technique does not lead to a complete characterization of the thermoelectric system without additional measurement. 
In order to extend impedance spectroscopy, we have investigated the measurement of the non-linear harmonic response of a Peltier device. 
The experiments are analyzed using an analytic model obtained by solving the heat equation in the frequency regime.
Our work shows that fitting the experimental response of the system in the harmonic regime can leads to a complete characterization of the thermoelectric properties without the need of additional measurement.

\end{abstract}

\maketitle

\section{Introduction}

Thermoelectric materials are ideal to develop energy harvesting and cooling systems without any mechanical moving part. 
The performance of a thermoelectric material or system depends on its figure of merit ($zT$). 
The higher the $zT$, the more efficient the material is and the better the system will perform.\cite{ioffe1957physics} 
As the $zT$ is a combination of three intrinsic properties ( $zT = \alpha^{2} T \sigma / \kappa$), complete characterization of thermoelectric materials or systems requires more than extracting the figure of merit. 
Individual measurement of thermal conductivity $\kappa$, Seebeck coefficient $\alpha$ and electrical conductivity $\sigma$ are then needed. 
This usually requires the use of several experimental setups with different boundary conditions in order to measure all physical properties, increasing the complexity for accurate thermoelectric measurements. 

The most used technique to study thermoelectric materials is the Harman method \cite{harman1958special}, which allows simultaneous measurement of both the $zT$ and the electrical resistivity. 
The difference between thermal and electrical time constants is at the core of this method in which the electrical response of the system is investigated in the time domain. 
However, the Seebeck coefficient and the thermal conductivity cannot be extracted from this technique. 

The existence of distinct thermal and electrical time constants can also be investigated in the frequency domain using impedance spectroscopy. Impedance spectroscopy is not a well known technique for thermoelectric measurement.
It is however, widely used in other scientific domains \cite{barsoukov2005impedance}.
In the framework of the linear response theory, the impedance spectroscopy measurements consist of measuring the in-phase and out-of-phase AC voltage responses to a small AC excitation current as function of the frequency. 
The impedance spectrum of a thermoelectric sample is characterized by a cutoff frequency corresponding to the time constant of the thermal diffusion across the system. 
At low frequency, below the thermal cutoff frequency, the response is the sum of the ohmic and thermoelectric responses. 
At high frequency, above the cutoff frequency, the response reduces to the pure ohmic contribution.
One particular advantage of impedance spectroscopy for thermoelectric measurement when compared to other methods is that there is no need to apply any thermal gradient during the measurement.

Recent works on impedance spectroscopy have shown that this is a powerful technique to characterize thermoelectric systems. 
The first attempt in 2002 by Dilhaire et al. \cite{dilhaire2002determination} showed that all physical properties (Seebeck coefficient, electric conductivity, thermal conductivity and figure of merit) can be extracted from one single linear impedance spectroscopy measurement on a PN thermoelectric junction if the thermal heat capacity have been previously measured. 
In their work an analytic model of the system, obtained by solving the heat equation in the harmonic regime with the quadripole method, was used to fit the impedance spectrum and to extract the desired physical properties. 

The impedance spectra can also be fitted with equivalent electrical circuits models (based on the low frequency asymptotic behavior).\cite{downey2007characterization, garcia2014low} 
The latter method allows a simpler analysis, however only few parameters can be extracted compare to the use of the analytic model (ohmic resistance, figure of merit and  thermal time constant are obtained). 

Impedance spectroscopy can also be used to investigate the presence of thermal contacts, heat loss by convection and heat transfer at the boundaries. \cite{casalegno2013frequency, de2014peltier, garcia2014impedance, Beltran-Pitarch2018Influence, apertet2017small}
This gives another advantage for impedance spectroscopy as it allows to study how the system will interact with the hot and cold sources which are known to impact the performance of the system. \cite{apertet2012optimal} 

Linear impedance spectroscopy is a powerful tool to characterize thermoelectric systems, an accuracy close to $1 \% $ \cite{de2011accurate} for the determination of the figure of merit ($zT$) and the ohmic resistance ($R$) can be obtained. 
Therefore, it can be used for a precise evaluation (with a better accuracy than the Harman method) of the figure of merit of both a single material and a Peltier module in a wide range of temperature.\cite{hasegawa2018temperature, mesalam2018towards, beltran-pitarch2018thermal}
However, the linear response cannot give a complete characterization of the system without additional measurement, it suffers from the same limitation as those of the Harman method. 
Thermal conductivity and Seebeck coefficient extraction are still lacking without additional measurement. 
To overcome this limitation, we propose the use of non-linearity which is directly related to irreversibility in the system. The non-linearity of a thermoelectric system during impedance spectroscopy measurement leads to the presence of additional harmonics in the response. \cite{moran2015frequency}

In this work, we describe the use of nonlinear impedance spectroscopy to separate each component of the $zT$ parameter and obtain a complete characterization of a thermoelectric system with a single experimental setup. 
The article is organized as follows. 
At first, the description of the origin of the DC response, the first and the second harmonic response of a thermoelectric system submitted to a periodic excitation is given. 
Then by solving the heat equation for the unileg system in the frequency domain with nonlinear effect and isothermal boundary conditions an analytic solution is derived, assuming a temperature independence of thermoelectric properties.
The equivalent circuit model is derived from asymptotic behavior of the analytic model and used to fit the out-of-phase component of the impedance spectra for both first and second harmonics. 
To go beyond the approximation of temperature independence of thermoelectric properties we calculate the amplitude of the second harmonic response in the quasi static regime with temperature dependence of physical properties.
It follows the presentation of experimental results which is divided in three parts.
At first we present an experimental comparison between the adiabatic measurement, which is the classic configuration and the isotherm configuration, needed for non-linear measurement.
In a second part, we show the measurement of the amplitude of the DC response of the system under harmonic excitation.
This measurement, when combined with the linear measurement allows a complete characterization of the system.
The last experimental result that we present here is the measurement of the second harmonic response as a function of the frequency, which gives additional information about the derivative of the impedance of the system as function of the temperature
Finally we present the verification of the Kramers-Kronig relations for the second harmonic impedance spectra to show the consistency of the measured spectra.

\section{Thermal and electrical response of a Peltier module under AC excitation}

\begin{figure}[!h]
\centering
\includegraphics[width=0.4\textwidth]{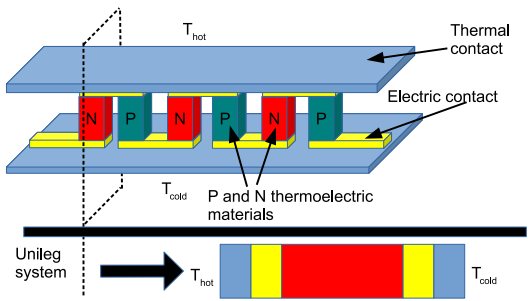}
\caption{Schematic view of a Peltier module and the associated one-dimensional model (unileg system). 
The unileg system is composed of one thermoelectric element, two electrical and two thermal contacts.}
\label{fig thermoelectric_device}
\end{figure}

In a general scheme, a Peltier module is composed of several junctions constructed with N and P materials electrically associated in series and thermally in parallel. 
The description of the module then reduces as a one-dimensional model equivalent to a single unileg system as shown in fig. \ref{fig thermoelectric_device}.
This unileg system is composed of a single thermoelectric element inserted between two electrical contacts and two thermal contacts.

For the sake of simplicity in this model, the thermal influence of the electrical contacts is neglected as they can be considered thin and highly conductive as compared to the thermoelectric materials.
We assume that material properties are temperature independent, neglecting the Thomson effect, as the variation of the temperature in the system remains small. 
Under these assumptions, the heat equation is expressed by:

\begin{align}
c \rho \frac{\partial T}{\partial t}(x) = \kappa \frac{\partial ^{2} T } {\partial x ^{2} }(x)    + \frac{J^{2}}{\sigma}
\label{eq}
\end{align}

Where $T$ is the temperature, $J$ the current density, $c$ is the massic heat capacity, $\rho$ is the volumic mass, $\kappa$ the thermal conductivity, and $\sigma$ the electrical conductivity. 
The heat equation is composed of three terms.
The first term (in the left member) corresponds to the heat capacity of the material.
The second term is the heat transported through conduction and the third term comes from the Joule effect. 
Two thermal sources are considered: the Joule effect that appears in the heat equation as $\frac{J^{2}}{\sigma}$, and the Peltier effect that appears in the boundary conditions.
The excitation of the system is driven by the AC current excitation express as Eq. (\ref{eq J}) with $\omega = 2 \pi f$ the angular frequency ($f$ is the frequency in Hz). 

\begin{align}
J(t) = \frac{J_{0}}{2}(e^{j \omega t} + e^{-j \omega t})     
\label{eq J} \\
T(x, \omega) = T_{DC} (x, \omega) + T_1 (x, \omega) e^{j \omega t} + \overline{ T_1} (x, \omega) e^{-j \omega t}
\nonumber \\
+ T_2 (x, \omega) e^{2j \omega t} + \overline{ T_2} (x, \omega) e^{-2j \omega t}
\label{eq T}
\end{align}

In order to solve the Eq. (\ref{eq}), under this conditions, it is possible to take advantage of the harmonic decomposition presented in Eq. (\ref{eq T}), where the DC term, the first and the second harmonics of the temperature, given by $T_{DC}(x, \omega)$, $T_{1}(x, \omega)$ and $T_{2}(x, \omega)$ respectively, are functions of the position $x$ and the angular frequency $\omega$.
$\overline{ T_1}(x, \omega)$ and $\overline{ T_2}(x, \omega)$ are the complex conjugate of $ T_1(x, \omega)$ and $ T_2(x, \omega)$ respectively.
The heat equation decomposition then gives:

\begin{align}
0 = \kappa \frac{\partial^{2} T_{DC} }{\partial x^{2}}(x, \omega)  + \frac{J_{0}^{2}}{2 \sigma} 
\nonumber \\
\rho c \omega T_{1}(x, \omega) = \kappa \frac{\partial^{2} T_{1} }{\partial x^{2}} (x, \omega)
\nonumber \\
2 \rho c \omega T_{2}(x, \omega) = \kappa \frac{\partial^{2} T_{2} }{\partial x^{2}} (x, \omega) + \frac{J_{0}^{2}}{2 \sigma} 
\label{eq app heat equation in freq}
\end{align}

Solution for $T_{DC}(x, \omega)$, $T_{1}(x, \omega)$ and $T_{2}(x, \omega)$ are given in Eq. (\ref{eq app t1 t2}).

\begin{align}
T_{DC}(x, \omega) = a_{DC} (\omega) + b_{DC}(\omega) x - \frac{J_{0}^{2} x^{2}}{4 \sigma \kappa}
\nonumber \\
T_{1} (x, \omega) = a_{1}(\omega) sh(q_{1}x) + b_{1}(\omega) ch(q_{1}x)             
\nonumber\\
T_{2} (x, \omega) = a_{2}(\omega) sh(q_{2}x) + b_{2}(\omega) ch(q_{2}x) + \frac{J_{0}^{2}}{4j \sigma \rho c \omega}               
\nonumber \\
q_{1} = \sqrt{\frac{j\rho c \omega}{\kappa}} 
,\quad
q_{2} = \sqrt{2}\sqrt{\frac{j\rho c \omega}{\kappa}}                
\label{eq app t1 t2}
\end{align}

Where $a_{DC}(\omega)$, $b_{DC}(\omega)$, $a_{1}(\omega)$, $b_{1}(\omega)$, $a_{2}(\omega)$ and $b_{2}(\omega)$ have the dimension of a temperature and are obtained by considering boundary conditions on the temperature.
$q_{1}$ and $q_{2}$ are thermal wavenumbers.
The Joule effect being non-linear, it depends on the square of the current density. 
Therefore, the associated thermal response in the harmonic regime will arise at twice the frequency of the current excitation.
The Peltier effect gives the quantity of the heat absorbed or released at the interface between two materials when a current crosses the interface: $Q = \Delta \alpha T I$. 
Where $\Delta \alpha$ is the difference of Seebeck coefficients between the two materials, $I$ is the electrical current and $T$ the temperature of the junction.
The Peltier effect need to be considered for the computation of the boundary condition on the temperature.

The voltage response of the system to the current excitation is composed of a pure electrical term (the ohmic response $V_{ohm} = RI$) and the thermoelectric response ($V_{te} = \Delta \alpha \Delta T$).
$\Delta T$ is the difference of temperature between the hot and cold junctions.
The overall voltage response is the product of the total impedance by the flowing current: $V_{te}+V_{ohm} =Z I$.
The harmonic decomposition of the impedance $Z$ is given by the DC, the first and the second-order terms given in Eq. (\ref{eq imp modeled}) as functions of the DC, first and second order temperature response, respectively.

\begin{align}
Z_{DC} =  \frac{\Delta \alpha \Delta  T_{DC} }{I}
\nonumber\\
Z_{1} =  R + \frac{\Delta \alpha \Delta T_{1}}{I}
\nonumber\\
Z_{2} =  \frac{\Delta \alpha \Delta  T_{2} }{I}
\label{eq imp modeled}
\end{align}

Previous works on Peltier module characterization with impedance spectroscopy were mostly performed in adiabatic conditions. 
\cite{downey2007characterization, de2011accurate, de2014peltier, garcia2014impedance} 
In this configuration, the sum of the thermal sources is zero.
The heating at one junction is canceled by the cooling at the other junction. 
The total heat brought into the system by the Peltier effect is zero allowing to consider steady state solution in adiabatic condition.
However, since the Joule effect is always positive, the heat transfer with a thermostat must be considered for a steady state solution of the temperature.
This is the case for a system under isotherm boundary condition.

\section{Linear impedance spectroscopy response}

\subsection{Linear adiabatic case.}

The linear response of a Peltier module under adiabatic condition has already been obtained (for example in \cite{garcia2016thermal}) based on the resolution of the heat equation in the frequency domain. 
In the linear approximation, only the thermal response at the excitation frequency ($f$) is considered ($V_{DC} = 0$ and $V_2 = 0$.
In this case, the impedance ($Z = Z_1$) is given by Eq.(\ref{eq z adia}).

\begin{align}
Z(\omega) = R \left(1+ \frac{zT}{F_{a} (\omega)}\right)
\nonumber \\
F_{a} (\omega) = \sqrt{\frac{j \omega }{ \omega _{t} }} coth  \left(   \sqrt{\frac{j \omega }{ \omega _{t} }}  \right)   + \frac{A}{2} \sqrt{\frac{j \omega }{ \omega _{c} }} th  \left(   \sqrt{\frac{j \omega }{ \omega _{c} }}  \right)  
\label{eq z adia}
\end{align}

The impedance depends on a dimensionless number ($F_{a}(\omega)$), the figure of merit ($zT$) and the electric resistance of the system ($R$).
The dimensionless number ($F_{a}(\omega)$) is a function of $\omega_{t}$, $\omega_{c}$ and $A$.
$\omega_{t}$ is the characteristic angular frequency related to the thermal diffusion in the thermoelectric material.
$\omega_{c}$ is the characteristic angular frequency related to the thermal diffusion in the thermal contacts.
$A$ is a dimensionless number related to the ratio of the thermal conductance in the contact divided by the thermal conductance in the thermoelectric material.

\begin{align}
zT = \frac{\sigma \alpha ^{2} T_{0} }{\kappa_{t}}
,\quad
R = \frac{L_{t}}{\sigma S}
,\quad
A = \frac{K_{c}}{K_{t}} =  \frac{\kappa_{c} L_{t}}{\kappa_{t} L_{c}}
\nonumber \\
\omega_{t} =  \frac{4 \kappa_{t}}{\rho_{t} c_{t} L_{t}^{2}}
,\quad
\omega_{c} =  \frac{\kappa_{c}}{\rho_{c} c_{c} L_{c}^{2}}
\label{eq Rztetc}
\end{align}

All five parameters given in Eq. (\ref{eq Rztetc}) are extracted when the experimental data are fitted with the analytic model. 
Note that $A = \frac{K_{c}}{K_{t}}$ plays an important role in the evaluation of the maximum power of a thermoelectric generator, as shown by Apertet et al. \cite{apertet2012optimal}.
The fitting parameters are functions of the thermoelectric system section ($S$), the thermoelectric element length ($L_{t}$), the thermal contact length ($L_{c}$) and thermal and electrical properties of both the thermal contacts and the thermoelectric materials.

\subsection{Linear isotherm case}

The Impedance spectrum of a Peltier module when one side is in direct contact with a thermostat and one side in adiabatic condition is given by Eq. (\ref{eq 1F no bottom contact}). 
This analytic solution is obtained when the thermal contact between the thermoelectric material and the thermostat is neglected. 

\begin{align}
Z_{1} =  R \left(  1 + \frac{zT }{F_{i}( \omega)}  \right) 
\nonumber\\
F_{i}( \omega ) =  \sqrt{\frac{j 4 \omega }{ \omega _{t} }} coth  \left(   \sqrt{\frac{j 4\omega }{ \omega _{t} }}  \right)   +  A \sqrt{\frac{j \omega }{ \omega _{c} }} th  \left(   \sqrt{\frac{j \omega }{ \omega _{c} }}  \right)      
\label{eq 1F no bottom contact}
\end{align}

Adiabatic and isotherm conditions lead to different spectra (a discusion of the difference between both thermal conditions can be found in \cite{hasegawa2018temperature}).
The transition from one condition to the other leads to a variation of the cut off frequency which is the results of the perturbation of thermal behavior of the system due to the contact with a thermostat.

At low frequencies, both adiabatic and isotherm measurements can be approximated by an equivalent circuit model obtained from the asymptotic behavior of the analytic model given in Eq. (\ref{eq z adia}) and (\ref{eq 1F no bottom contact}). 
In table \ref{tab tau}, we compare the characteristic time in both cases. 
With a $RC$ model, only three quantities can be extracted from the fit, the ohmic resistance $R$, the figure of merit $zT$ and a characteristic time $\tau$. 
The characteristic time ($\tau$) of the $RC$ model is a function of $\omega _{t}$, $\omega _{c}$ and $A$ from the analytic model. 
The reduction from the analytic to the RC model is obtained by computing the asymptotic behavior of the analytic model at low frequency.
In isotherm condition the characteristic time is identical to the characteristic time of an equivalent system in adiabatic condition with a thermoelectric length equal to the double of the thermoelectric length in the system in adiabatic condition ($\tau _{iso} (L_t) = \tau _{adia} (2 L_t)$).

\begin{table}[!h]
\centering
\begin{tabular}{ c | c | c }
    & RC model & Heat equation \\
   Impedance ($Z$) & 
   \begin{large}
   $R \left( 1 + \frac{  zT} {  1 + j \omega \tau }\right)$ 
   \end{large}
   &
   solution
\\\hline
   Adiabatic & 
   \begin{large}
   $\tau _{adia} = \frac{1}{3 \omega _{t}}+\frac{A}{2 \omega _{c}}  $
   \end{large}
   &
   Eq. (\ref{eq z adia})
\\\hline
   Isotherm & 
   \begin{large}
   $\tau _{iso}  = \frac{4}{3 \omega _{t}}+ \frac{A}{\omega _{c}} $
   \end{large}
   &
   Eq. (\ref{eq 1F no bottom contact})
\end{tabular}
\caption{Low-frequency asymptotic behavior of the impedance for the adiabatic and the isotherm cases.}
\label{tab tau} 
\end{table}

Comparison between linear adiabatic and isotherm case allows to verify the consistency of our model and the possibility to measure both the linear and non-linear response under the same experimental condition.

\section{Non linear response}

Non-linearity comes from two effects: the Joule effect and the Peltier effect. 
The Peltier effect induces heating or cooling given by $ Q = \Delta \alpha I T$. 
In the linear approximation, the temperature of the junction is approximated by the thermostat temperature $T_{0}$ which gives a linear Peltier response of $ Q_{1} = \Delta \alpha I T_{0}$. 
The second-order Peltier effect is given by $ Q_{2} = \Delta \alpha I T_{1}(L_t)$, where $T_{1}(L_t)$ is the first-order temperature response at the junction.

To study the effect of non-linearity, we consider the case of a unileg system under isotherm condition where the bottom contact is neglected (no thermal resistance between the Peltier module and the thermostat). 
The second-order thermal response is composed of Joule ($\Delta T_{2J}$) and second-order Peltier ($\Delta T_{2J}$).
The analytic solution of both these effects is given in Eq. (\ref{eq non linear T}). 
Both nonlinear Peltier and Joule responses depend on $F( \omega )$ and $G( \omega )$ functions as defined in Eq. (\ref{eq non linear T}).

\begin{align}
\Delta T_{2P} =  \frac{\alpha ^{2} I^{2} T_{0} }{2 F( \omega) F(2 \omega) K_{t}^{2}} 
\nonumber \\
\Delta T_{2J} =  \frac{ RI ^{2}}{K _{t}}  \frac{G(2 \omega )  }{2 F(2 \omega )}    
\nonumber \\
F( \omega ) = \sqrt{\frac{j 4 \omega }{ \omega _{t} }} coth  \left(   \sqrt{\frac{j 4\omega }{ \omega _{t} }}  \right)        + A \sqrt{\frac{j \omega }{ \omega _{c} }} th  \left(   \sqrt{\frac{j \omega }{ \omega _{c} }}  \right)   
\nonumber \\
G ( \omega ) = \frac{ch  \left(   \sqrt{\frac{j 4\omega }{ \omega _{t} }}  \right) -1}{ \sqrt{\frac{j 4\omega }{ \omega _{t} }}  sh  \left(   \sqrt{\frac{j 4\omega }{ \omega _{t} }}  \right) }      \quad
\label{eq non linear T}
\end{align}

The second-order impedance is given in Eq. (\ref{eq non linear Z}) following the same definition for $F( \omega )$ and $G( \omega )$.

\begin{align}
Z_{2} = \frac{\alpha \left(  \Delta T_{2P}  +    \Delta T_{2J}  \right)}{I}
\nonumber\\
Z_{2} =  \frac{ \alpha RI  }{2 K _{t}F(2 \omega )} \left(  G(2 \omega  ) + \frac{  zT}{F(\omega)} \right)
\label{eq non linear Z}
\end{align}

The high-frequency limit of the second-order impedance is zero. 
The low-frequency response is the sum of the stationary solution for the Joule and the second-order Peltier effects ($ \alpha RI  (1 + 2zT)/(4 K_{t}) $).
The analytic solution given in Eq. (\ref{eq non linear Z}) can be approximated by an equivalent circuit model at low frequencies based on the low frequency asymptotic behavior of the analytic model. 
The equivalent RC model then reads :

\begin{align}
Z_{2} = 
\frac{
\left( 
1 + 2zT
\right)  
\alpha RI  }
{ 
 4 K _{t} 
\left( 
1 +  
j \omega \tau _{2}
\right)
} 
\label{eq non linear Z RC}
\end{align}

The non-linear effects also appears as a DC response when the system is under AC excitation.
The DC response ($Z_{DC}$ in Eq. \ref{eq z0}) is similar to the 2f response ($Z_2$ in Eq. \ref{eq non linear Z}) with one contribution from Joule effect and one contribution from the non linear Peltier effect.
The main difference is that the amplitude the Joule part of the response will be independent of the excitation frequency.
In this model the high frequency DC impedance is given by equation (\ref{eq z0 hf}) as the results of the Joule effect alone.

\begin{align}
Z_{DC} =  \frac{ \alpha RI  }{2 K _{t}} \left(  \frac{1}{2} + Re \left[ \frac{zT}{F(\omega)} \right] \right)
\label{eq z0} \\
Z_{DC \, f \rightarrow \infty} =  \frac{ \alpha RI  }{4 K _{t}}  
\label{eq z0 hf}
\end{align}

\begin{figure}[!h]
\centering
\includegraphics[width=0.4\textwidth]{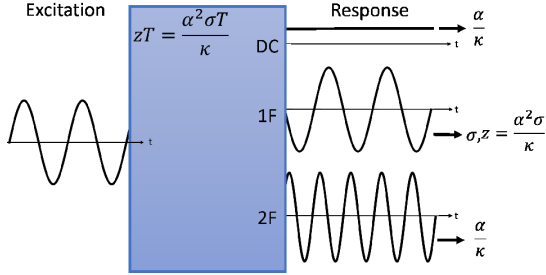}
\caption{
Linear and nonlinear contributions of the thermoelectric response to an harmonic excitation.
}
\label{fig 1f2f}
\end{figure}

The interest in the non linear measurement comes from the ability to extract additional informations about the system if the DC and the second harmonic are measured.
In Fig. \ref{fig 1f2f}, the quantities extracted by fitting the experimental result with the model are shown.
This method gives a complete characterization of the thermoelectric system since all thermoelectric properties ($\alpha$, $\kappa$, $\sigma$ and $zT$) can be extracted.
However, this result is invalid if the temperature dependence of material properties cannot be neglected as demonstrated in the following section.

\section{Quasi static response with temperature dependent properties.}

In order to evaluate the effect of temperature dependence of material properties, the second harmonic of the impedance is computed in the quasi static regime.
In general the influence of temperature dependence of physical properties are non longer negligible in the non-linear response.
The current excitation is given by Eq. (\ref{eq I QS}) which is equivalent to the current density given in Eq. (\ref{eq J}).

\begin{align}
I(t) = I_{1}cos ( \omega t)     
\label{eq I QS}
\end{align}

Besides, since in the approximation of non-linear response, only the DC, the first and the second harmonic responses are considered, the temperature and the voltage responses are given by Eq. (\ref{eq_T_V_QS}) in the quasi static regime. 

\begin{align}
T(t) = T_{0} + T_{DC}(x) + T_{1}(x) cos (\omega t) + T_{2}(x) cos (2 \omega t)    
\nonumber \\
V(t) =  Z_{DC} I + Z_{1} I cos (\omega t) + Z_{2} I cos (2 \omega t)  
\quad 
\label{eq_T_V_QS}
\end{align}

The solution for $ T_{DC}(x)$, $ T_{1}(x)$ and $ T_{2}(x)$ can be computed by solving the heat equation in the quasi static regime.
Thermoelectric properties are temperature dependent and in the approximation of small temperature variations the properties of the system are given by Eq. (\ref{eq_Taylor_props}) with $R'(T_{0})  =  \frac{\partial R}{\partial T}(T_{0}) $, $\alpha' (T_{0}) = \frac{\partial \alpha}{\partial T} (T_{0}) $ and $K' (T_{0}) =  \frac{\partial K}{\partial T} (T_{0})$

\begin{align}
R(t) = R(T_{0}) + (T(t) - T_{0}) R' (T_{0}) 
\nonumber \\
\alpha (t) = \alpha(T_{0}) + (T(t) - T_{0}) \alpha' (T_{0}) 
\nonumber \\
K(t) = K(T_{0}) + (T(t) - T_{0})  K' (T_{0})
\label{eq_Taylor_props}
\end{align}

The voltage response is obtained from the temperature response (Eq. (\ref{eq_sol_v})).

\begin{align}
V = R(T(t)) I(t) + \alpha (T(L_{t})) \Delta T(t)
\label{eq_sol_v}
\end{align}

The linear voltage response is given by $V_{lin} = Z_{1} I$, where this notation takes into account the linear ohmic and thermoelectric responses.
The non-linear effect arises from two different sources: the variation of the thermoelectric properties with the temperature and the Joule heating.
The variation of the thermoelectric properties with the temperature can be described for small temperature variations by a first order Taylor expansion of the impedance $Z_{1}$.
The Joule effect describes a heating source of $Q_{J} = R I^{2}$ in a system with an electrical resistance $R$. 
In a system with an impedance $Z$, this heating source is given by the generalized Joule effect $Q = Z I^{2}$.

\begin{align}
V = Z_{1} I + Z_{1}' \langle \Delta T \rangle I + \frac{\alpha Z_{1} I ^{2} }{2 K},
\label{eq_sol_v_static}
\end{align}

where $Z_{1}$ is the linear impedance and $Z_{1}'$ is the partial derivative of the linear impedance with respect to the temperature at the temperature $T_{0}$ of the thermostat.
The total impedance in static condition is given by the sum of three terms: the linear part, the first-order Taylor expansion and the generalized Joule effect.

\begin{align}
Z_{1} = R + \frac{\alpha ^{2} T_{0}}{K}
\nonumber \\
Z_{1}' = R' + \frac{\alpha ^{2}}{K} + \frac{2 \alpha ' \alpha T_{0} }{K} - \frac{\alpha ^{2} T_{0} K'}{K^{2}}
\label{eq_zdz}
\end{align}

The temperature difference $\Delta T$ can be approximated by the contribution of the Peltier effect ($ \Delta T = \frac{ \alpha T I }{K}$) if we are close to the linear regime.
The average temperature difference $\langle \Delta T \rangle $ equals half the temperature difference $ \frac{ \alpha T I ^{2} }{2 K}$. 
The Eq. (\ref{eq_sol_v_static2}) gives the voltage response in the static regime, the quasi static regime is obtained when an alternative current given by Eq. (\ref{eq I QS}) is injected in Eq. (\ref{eq_sol_v_static2}).

\begin{align}
V = Z_{1} I + Z_{1}' \frac{ \alpha T I ^{2} }{2 K} + \frac{\alpha Z_{1}I ^{2} }{2 K}
\label{eq_sol_v_static2}
\end{align}

The second harmonic voltage in then given by Eq. (\ref{eq_sol_v2}), it is proportional to the derivative of $Z_1 T$ with respect to the temperature.

\begin{align}
V_{2} = \frac{\alpha I ^{2} }{4 K} \left( Z_{1} + Z_{1}' T_{0} \right) 
= 
\frac{\alpha I ^{2} }{4 K} \frac{ \partial \left( Z_{1} T \right) }{\partial T}
\label{eq_sol_v2}
\end{align}

The second harmonic voltage written as a function of the systems properties ($R$, $K$, $\alpha$) and their derivative with respect to the temperature ($R '$, $K '$, $\alpha '$) is given in Eq. (\ref{eq_sol_v2_2}) where $z = \frac{\alpha ^2}{R K}$.
Eq. (\ref{eq_sol_v2_2}) gives the low-frequency limits ($\omega = 0$) of Eq. (\ref{eq non linear Z RC}) when $R'= 0 $, $K'= 0 $ and $\alpha '= 0 $ (temperature independent properties).

\begin{align}
V_{2} = 
\frac{\alpha I ^{2} R }{4 K} 
(
1 + 2 zT_{0} + \frac{R'}{R} T_{0} + 
\nonumber \\
2 \frac{\alpha '}{\alpha} z T_{0} ^{2} - \frac{K'}{K} z T_{0} ^{2}
)
\label{eq_sol_v2_2}
\end{align}

For conventional materials, $\frac{R'}{R}$, $\frac{K'}{K}$ and $\frac{\alpha '}{\alpha}$ are on the order of magnitude of one percent \cite{yoo2017impedance}.
Therefore, in the second harmonic response, the influence of temperature dependence of material properties cannot be neglected.
From equation (\ref{eq_sol_v2_2}) we can deduce that if $\frac{K'}{K}T_0 \ll 1 $, $\frac{R'}{R}T_0 \ll 1  $ and $\frac{\alpha '}{\alpha}T_0 \ll 1  $ the system can be approximate as a system with temperature independent properties.

\section{Experimental results}

Analysis of both linear and nonlinear effects was performed on a commercial mini Peltier module (03201-9U30-08RA) from Custom Thermoelectric.
Our non-linear model was used to go beyond the results obtain based on the linear impedance spectroscopy measurement. 
The DC measurement was performed with an oscilloscope and a Keithley 6221.
The impedance response at $1f$ and $2f$ of a thermoelectric module was obtained with a DSP lock-in amplifier (model 7230 from AMETEK) with an AC current source of $15$ mA$_{RMS}$. 
This apparatus allows the simultaneous measurement of the first and second harmonics.

\subsection{First harmonic response}

\begin{figure}[!h]
\centering
\includegraphics[width=0.4\textwidth]{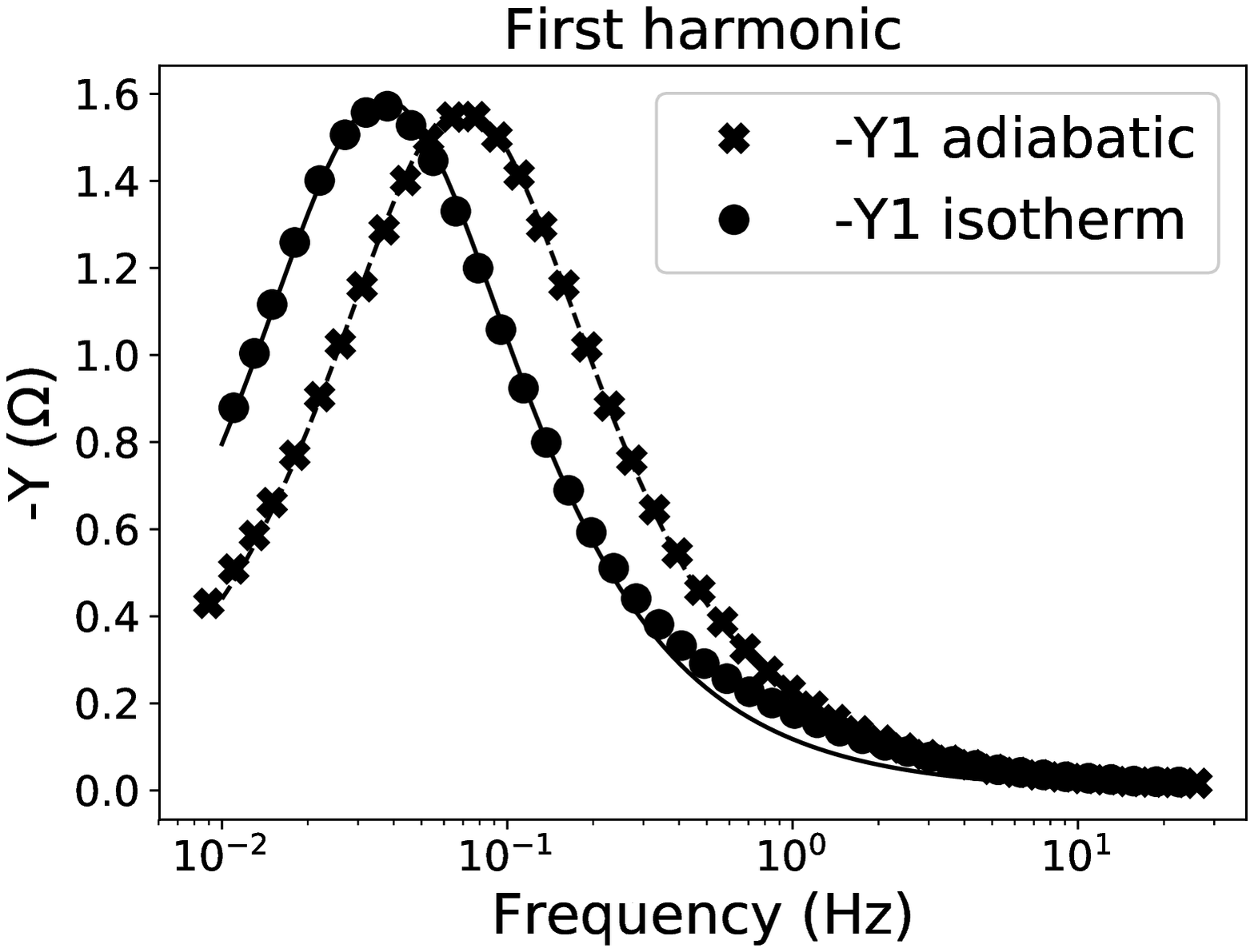}
\caption{
The plot of the measured imaginary part of the impedance as a function of the frequency. 
The adiabatic case is plotted in blue and the isotherm case in red. 
Both measurements were fitted with an equivalent circuit model and their associated fit is plotted as straight lines. 
A frequency shift is clearly visible with a lower characteristic frequency for the isotherm case.
}
\label{fig comparison 1f}
\end{figure}

The imaginary part of the impedance ($Y = Im(Z)$) is plotted in Fig. \ref{fig comparison 1f} in both the adiabatic and the isotherm cases. 
The adiabatic case is obtained when the Peltier is suspended by its contact wire.
In isotherm condition the Peltier module is in contact with a steel plate that act as a thermostat.
A good thermal contact is obtain with thermal grease.
A frequency shift is expected with a higher characteristic frequency for the adiabatic case. 
Indeed, this behavior is observed with a characteristic frequency of $70$ mHz in the adiabatic case and of $37$ mHz in the isotherm case. 
However, the adiabatic characteristic time multiplied by two ($2 \times 14.5 = 29s$) is higher than the characteristic time in the isotherm case ($27s$).
This is not expected according to the theoretical values of the characteristic times given in Table \ref{tab tau}. 

This inconstancy finds its origin in the asymmetry of the Peltier module. 
When measured in an adiabatic condition, the measured characteristic frequency is a combination of both top and bottom thermal contacts and of the thermoelectric material characteristic frequencies. 
However, under isotherm condition, only the thermoelectric material and the thermal contact which is not in contact with the thermostat influence the characteristic frequency. 
When the system is measured under isotherm condition with the second thermal contact as the contact between the system and the thermostat, a characteristic frequency of $33$ mHz is obtained corresponding to a characteristic time of $30s$. 
These results evidence the key role of the asymmetry of the system.

The first harmonic measurement gives an ohmic resistance of $4.9 \Omega$ and a figure of merit of $0.72$ with a difference of $1\%$ and $3\%$ respectively between the adiabatic and isotherm case. 
These results are consistent with the properties of the measured system with a significant increase ($+13 \%$) of the electrical resistivity (from $4.3 \Omega$ to $4.9 \Omega$) as a function of time which is due to the aging of the system.

\subsection{Non-linear high-frequency DC response}

As the origin of the high frequency DC response is the Joule effect alone, it remains unchanged even in the case where the approximation of temperature invariant properties is no longer valid.
By associating the figure of merit and the ohmic resistance measurement from the linear impedance spectroscopy with the measurement of $\frac{\alpha R }{4 K} $ from DC voltage measurement as a function of the AC current amplitude at high frequency, all thermoelectric properties can be extracted.
The model for the high-frequency DC impedance response is given in Eq. (\ref{eq z0 hf}).
The DC voltage is deduce from the DC impedance with $V_{DC} = Z_{DC} I$

\begin{figure}[!h]
\centering
\includegraphics[width=0.4\textwidth]{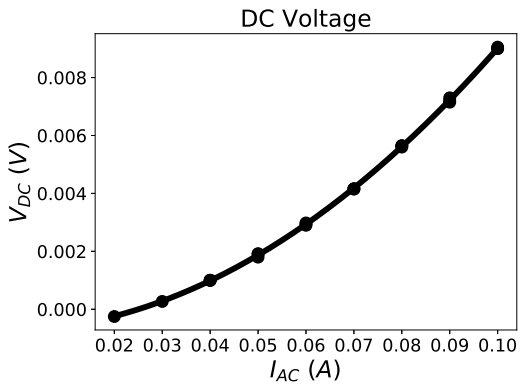}
\caption{
DC voltage response as a function of the AC current amplitude and the second-order polynomial fit.
}
\label{fig dc}
\end{figure}

A 1kHz AC current excitation in the range of $20$ mA to $100$ mA is obtain with a Keithley 6221. A simple low pass filter is used to isolate the DC voltage which is measured. The fit of the measured DC voltage as a function of the AC current amplitude by a second order polynome (shown in Fig \ref{fig dc}) gives the value of $\frac{\alpha R }{4 K} = 0.91 \Omega A^{-1}$

From the first harmonic response and the DC response, the Seebeck coefficient of the system is $\alpha_{Peltier} =16$ $mV.K^{-1}$ and the thermal conductance is $K_{t} = 21$ $mWK^{-1}$. 
The Peltier device is composed of 64 thermoelectric legs. Individual legs are $1.1$ mm in height and have a section of $0.17$ $mm^{2}$.
By taking into account the geometry of the device, the properties of the average material that compose the thermoelectric legs can be computed.
The average calculated Seebeck coefficient from experimental results of P and the N material is $\alpha = 250$ $\mu V K^{-1}$.
The average thermal conductivity and the average electrical conductivity are $\kappa = 2.1$ $W K^{-1} m^{-1}$ and $\sigma = 8.2 $ $ 10^{4} S.m^{-1}$, respectively. 
The measured thermoelectric properties are similar with measurement (\cite{hasegawa2018temperature}) of this material ($Bi_2Te_3$) at room temperature.

\subsection{Second harmonic response}

\begin{figure}[!h]
\centering
\includegraphics[width=0.4\textwidth]{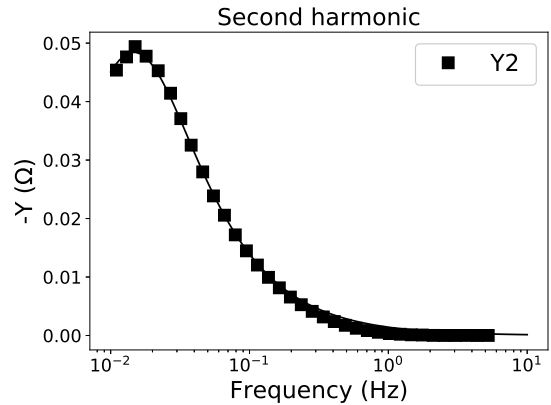}
\caption{
Imaginary part of the second-order impedance as a function of the frequency and the fit with an equivalent circuit model.
}
\label{fig 2f}
\end{figure}

The experimental results of the second harmonic response and the associated fit are plotted in Fig. (\ref{fig 2f}). 
The fit of the second-order response gives a characteristic frequency of $15$ $mHz$ and a low frequency limit for the second harmonic impedance of $ Z_{2 \, f \rightarrow 0} = 97$ $m \Omega$. 
$ Z_{2 \, f \rightarrow 0}$ is the quasi static impedance defined as $V_2/I$ with $V_2$ given in Eq. (\ref{eq_sol_v2_2}).

The second harmonic voltage given in Eq. (\ref{eq_sol_v2_2}) is written as a sum of five terms.
The term $\frac{\alpha I^2R}{4K}$ is the results of the Joule effect.
The term $\frac{\alpha I^2RzT_0}{K}$ is the results of the non linear Peltier effect.
The other terms are the results of the temperature dependence of thermoelectric properties ($K$, $R$, $\alpha$).
From our experimental results, we deduce that the Joule effect gives $20 \%$ of the second harmonic response and the temperature dependence of the properties gives $47 \%$ of the second harmonic response.
This shows that the Joule effect is not the preponderant non-linear effect on the second harmonic response and the influence of the temperature dependence of the system properties plays an important role in the second harmonic response described in section V.

\section{Kramers-Kronig relations}

Kramers-Kronig relations are used to assess the consistency of experimental data of impedance spectroscopy measurements and to insure that the measured spectra is the consequence of the excitation. \cite{boukamp1993practical, agarwal1995application} 
Kramers-Kronig relations between the real part ($X$) and the imaginary part ($Y$) of the impedance are given in Eq. (\ref{eq kk}).

\begin{align}
X(\omega) - X(0) =
\frac{2 \omega}{\pi}
\int_{0}^{\infty}
\frac{
( \omega / x ) Y(x) - Y(\omega)
}
{
x^{2} - \omega ^{2}
}
dx 
\nonumber \\
Y(\omega) =
\frac{2 \omega}{\pi}
\int_{0}^{\infty} 
\frac{
X(x) - X(\omega)
}
{
x^{2} - \omega ^{2}
}
dx
\label{eq kk}
\end{align}

This analysis is usually performed on linear systems, however it can be extended to non-linear systems. \cite{peiponen2004kramers} 
Each harmonic of the response should respect Kramer Kronig relations.
In our case we have verified these relations for the first and the second-order terms of the impedance of the system.

In order to obtain a good numerical integration from zero to the infinity, a wide frequency range is required.
Experimental data are only available on a restricted frequency range. 
Therefore, experimental data were extended with data calculated from the fit of the measured data with an equivalent circuit model.
From experimental and extended data, a numerical integration is used to calculate a new set of $X$ and $Y$ with Kramers-Kronig relations. 

\begin{figure}[!h]
\centering
\includegraphics[width=0.4\textwidth]{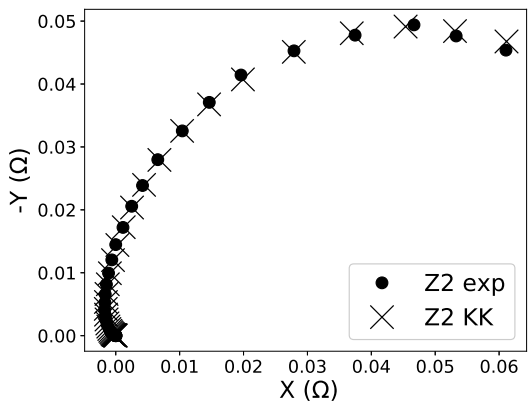}
\caption{
Nyquist plot of the second order impedance for the measured data (circles) and the calculated data from the Kramers-Kronig relation (crosses).
}
\label{fig kk}
\end{figure}

In Fig. \ref{fig kk} experimental and calculated values of the impedance in a Nyquist plot are shown.
The normalized deviation between experimental and calculated values for the first and second harmonic is lower than $4 \%$.
This shows that the relation between the real and imaginary part of first and second order impedance of a thermoelectric system follows Kramers-Kronig relations.
The good agreement of our measured spectra with respect to the Kramers-Kronig relations allows to validate the consistency of the measured spectra.

\section{Conclusion}

Impedance spectroscopy is a promising technique to characterize thermoelectric systems. 
By separating the thermal and the electrical responses in the frequency domain, impedance spectroscopy measurement provides all key parameters to characterize thermoelectric systems. 
The electric resistance and the $zT$ are easily extracted with linear impedance spectroscopy.
However, a complete characterization is not possible without additional measurement. 
In this work, we have investigated the nonlinear response by taking into account the Joule heating and nonlinear Peltier effects. 
Nonlinear effect implies the appearance of higher harmonics in the impedance response. 
In our example, we studied the DC response, the first and the second harmonic responses. 
The second harmonic response is the sum of different non-linear effects.
Our results show that the Joule effect is not the preponderant non-linear effect, as it only contributes to $20 \%$ of the total second harmonic response. 

The association of the DC and the linear impedance measurements allows a complete characterization of the thermoelectric system.
The results of our work show the interest to consider the nonlinear response of the system during impedance spectroscopy measurement, and the use of it as a standalone technique for complete characterization ($zT, \, \alpha, \, \kappa, \, \sigma$) of thermoelectric system.

\begin{acknowledgments}
The authors thank ANRT (CIFRE) for the funding of doctoral studies by E. Thiebaut, ST Microelectronics TOURS for their support and Dr. Chun-Yul Yoo for fruitful discussions.
\end{acknowledgments}

\end{document}